\pdfoutput=1
\documentclass[conference]{IEEEtran}
\IEEEoverridecommandlockouts
\usepackage{subcaption}
\usepackage{cite}
\usepackage{amsmath,amssymb,amsfonts}
\usepackage{algorithmic}
\usepackage{graphicx}
\usepackage{textcomp}
\usepackage{xcolor}

\def\BibTeX{{\rm B\kern-.05em{\sc i\kern-.025em b}\kern-.08em
    T\kern-.1667em\lower.7ex\hbox{E}\kern-.125emX}}
\begin{document}

\title{Energy Storage System (ESS) for Compensating Unbalanced Multi-microgrids Using Modified Reverse Droop Control\\
}
\author{\IEEEauthorblockN{Mehmet Emin Akdogan and Sara Ahmed}
\IEEEauthorblockA{\textit{Electrical and Computer Engineering} \\
\textit{The University of Texas at San Antonio}\\
San Antonio, Texas \\
m.eminakdogan@gmail.com, sara.ahmed@utsa.edu}
}

\maketitle
\begin{abstract}
The connection of single-phase microgrids (MG) and loads to three-phase MGs creates power quality problems such as unbalanced voltage and voltage rise at the point of common coupling (PCC) of the MGs.  In this paper, a modified reverse droop control (MRDC) scheme in the Energy Storage System (ESS) is proposed to improve the three-phase PCC voltage quality in multi-microgrids (MMG). The MRDC consists of a reactive power compensator (RPC) and a voltage compensator. The controller regulates the reactive power and voltage unbalance of the MMG by using the reactive power produced by the ESS. The effectiveness of this proposed scheme is verified in real-time simulation using the Opal-RT OP5600 real-time simulator. The voltage unbalance factor (VUF) at the PCC is decreased from 3.6 percent to 0.25 percent, while the reactive power is reduced significantly at the single-phase load.

\end{abstract}

\begin{IEEEkeywords}
Unbalanced voltage compensation, reactive power compensation, reverse droop control, distributed generations, PV islanded, energy storage system, voltage controlled inverters, multi-microgrids, power quality.
\end{IEEEkeywords}

\section{Introduction}
In recent years, a significant amount of renewable distributed energy resources (DERs) has been integrated into both  transmission  and  distribution  power  systems. A micro grid (MG) is a low-voltage distribution network that integrates multiple DERs, ESSs and dispersed loads monitored and controlled by an energy management system (EMS) to reduce the impact of high distributed generation (DG) penetration on power system operation and improve the service reliability. It can operate both in main grid connected mode and autonomous island mode \cite{HMG2,RPCSP3,ESShybrid,KHAN2021,apec,akdogan}.

DERs in microgrids, particularly residential rooftop photovoltaic (PV), have provided a clean and cost-effective solution for remote areas (rural and mountainous areas) with no access to the utility grid. Therefore, islanded multi microgrids (MMG) that include three and single phase microgrids have increased significantly in recent decades \cite{IET,MMG,HMG1}. The high penetration of single-phase loads and single-phase PV-DGs in three phase system have caused several challenges like voltage unbalance, voltage increase and unequal power flows in MMGs. As a result, the unbalanced MMGs will decrease the lifespan machines, increase power losses, and reduce hosting capacity\cite{mypaper,IET}. Therefore, different control methods for unbalanced MMGs and low voltage distribution networks (LVDNs) have been discussed in the literature to overcome voltage quality problems\cite{MMG,apf,UPQ,af,HMG1,sequence,03,Multiobjective,RPCSP1,RPCSP2, ESS1,ESS2,chinmay,Droop1,Droop2,Droop3,Droop4,Droop5,ESSdroop1,ESSdroop2,mypaper}.

Unbalanced voltage active compensation using power quality conditioners, such as static var compensators (SVC), static synchronous compensators (STATCOMs), active power filters (APFs) and unified power quality conditioners has been used for years. \cite{apf,UPQ,af}. However, the power quality conditioners significantly increases the initial and operational costs of the system. Active compensation by controlling the DG inverters was adapted \cite{HMG1, sequence,MMG,03,Multiobjective}.

For example, a hierarchical event-based distributed control method \cite{HMG1} and a sequence-based current controller with reactive power compensation \cite{sequence} have been proposed to enhance the voltage unbalance at the PCC and the DG terminals in the MMGs. Robust and compound control strategies based on quasi proportional resonant for voltage control compensation are implemented in an islanded MMG \cite{MMG,03}. In addition, the authors in \cite{Multiobjective} proposed an optimal multi-objective control algorithm applied to the secondary level to mitigate the voltage unbalance and optimally regulate the power flow  among the system phases in AC-DC hybrid MGs. 

\begin{figure}
\centerline{\includegraphics[width=1.\linewidth]{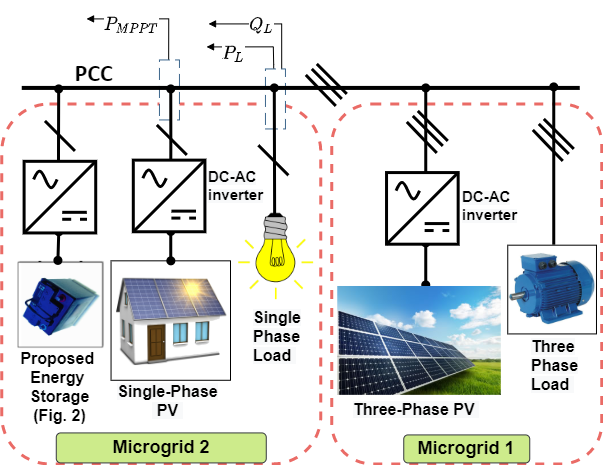}}
\caption{A typical unbalanced multimicrogrid system with PV and ESS systems}  
\vspace{-11pt}
\label{Simplified}
\end{figure}

Some studies have been proposed using the surplus power capability of the single-phase DG inverters to control the reactive power exchange for voltage regulation in unbalanced MMGs \cite{RPCSP1,RPCSP2}.  The method in \cite{RPCSP1} consists of two groups of PV inverters in the form of distributed delta and wye connections. The delta connections are used to control the reactive power exchange between phases to compensate the negative and zero sequence components of the PCC voltage, while the wye connections between the phases to neutral are utilized to manage the reactive power sharing between PV inverters. In addition, Karush–Kuhn–Tucker (KKT) optimization method of the single-phase DGs smart inverters is presented in \cite{RPCSP2} for negative and zero-sequence current compensation to enhance the voltage unbalance in the MMGs. However, the reactive power compensation controlled by the single-phase DG inverters may reduce the DGs’ lifetime \cite{ESS1}.

Using a single-phase ESS connected to a single-phase PV inverter is also another solution instead of power curtailment, which increases the cost of the power generation \cite{ESS1, ESS2,chinmay}. The authors presented a voltage regulation strategy in the LVDNs using the distributed battery energy storage (BES) systems in the rooftop PV systems for mitigating voltage unbalance of the network with the high penetration of PV systems \cite{ESS1}. In addition, the objective of \cite{ESS2} is to develop an unbalanced voltage mitigating method in a distributed energy storage (DES) integrated with a four-quadrant converter. The DES inject real and reactive power to the network to reduce voltage-unbalance and losses of the network. However, regulating unbalanced voltage in islanded operation of the MMGs is not considered in \cite{ESS1, ESS2}.


DG inverters in the primary control of the MG system can be operated in either voltage control mode (VCM) or current control mode (CCM). While conventional droop control has been implemented in VCM inverters for primary regulation of frequency and voltage, reverse droop control have been adopted by CCM inverters to provide coordinated control of the active and reactive power for current regulation. \cite{Droop1, Droop2, Droop3, Droop4,Droop5}.
In \cite{Droop4}, a stochastic mixed-integer nonlinear programming (MINLP) model as an extension of an optimal power flow formulation is presented for voltage regulation and reducing frequency deviation in unbalanced three-phase droop-based microgrid. In addition, droop per phase control and secondary control algorithms are proposed in \cite{Droop3} for regulating unbalanced three-phase four-wire MGs. Modified reactive droop and reverse droop control schemes in \cite{Droop5} are developed based on the coordination of VCM and CCM converters respectively, for  reactive power and harmonic load compensation.

Droop control methods are implemented in battery energy storage system (BESS) for voltage regulation in low voltage (LV) microgrid with high penetration of PV generation \cite{ESSdroop1,ESSdroop2}. A fuzzy logic based droop control strategy in \cite{ESSdroop1}  mitigates over/under voltage situations during peak PV generation or low power consumption respectively. In addition, an event-triggered and a leader following distributed voltage control strategies among adjacent BESSs have been proposed to regulate the bus voltages in LVDN interconnected with large-scale solar PVs \cite{ESSdroop2}. 
 
No methods have been proposed to use the reverse droop control to reduce the unbalanced voltage at the PCC by regulating active and reactive power in an ESS. Therefore, this paper proposes a modified reverse droop control (MRDC) for the ESS. This proposed MRDC consists of a reactive power compensator (RPC) that generates the active power reference of the ESS to enhance the power quality.  It also includes a voltage compensator to balance the PCC voltage of the three-phase PV within the allowed limits in the unbalanced MGs.  The main contributions of the paper are as follows:  

\begin{itemize}
\item A new control strategy for the ESS based on reverse droop control to compensate for the voltage at the PCC terminal. To the best of the author’s knowledge, reverse droop controllers have not been implemented to an ESS or single phase inverter before.	
\item Modified reverse droop control to act as a RPC without needing any central (secondary) controller.	
\end{itemize}

The paper is organized as follows: the system structure and proposed control scheme, including the RPC and the voltage compensator are explained in Section II. Simulation results in Matlab/Simulink and experimental results in real time are presented in Section III and finally, section IV concludes this paper.

\begin{figure*}[htbp]
\centering
\includegraphics[width=0.85\linewidth]{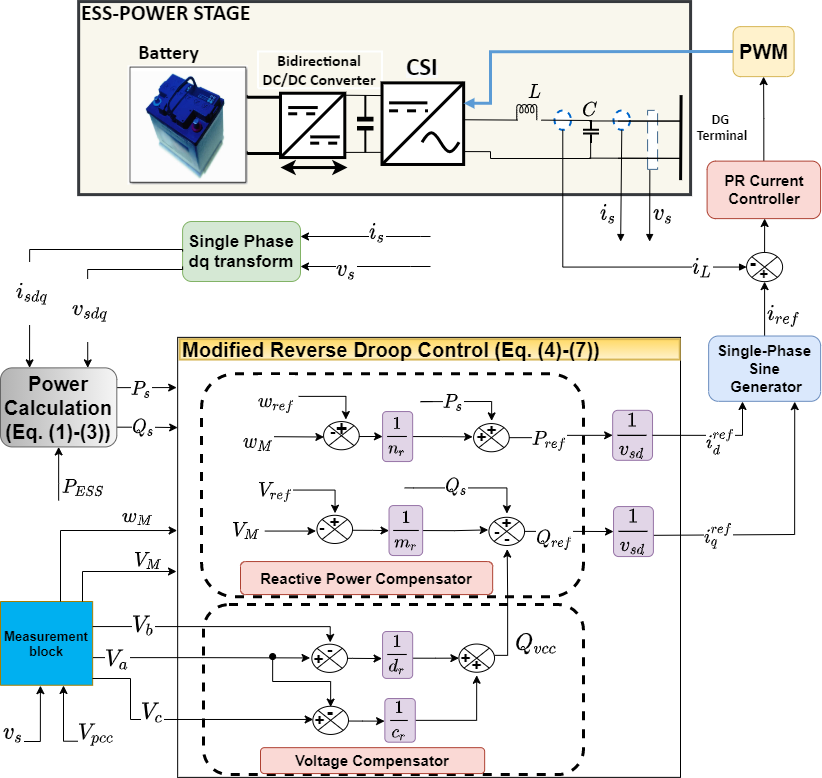}
\caption{Detailed block diagram of proposed control scheme for the ith ESS.}
\label{Detailed}
\vspace{-11pt}
\end{figure*}

\section{the system structure and  proposed control scheme }
An unbalanced multimicrogrid with the proposed ESS  connected in parallel to the single-phase PV inverter and a single phase load is  shown in Fig.~\ref{Simplified}. The system consists of two MGs where microgrid 1 is a three-phase PV-MG in parallel with microgrid 2, a single-phase PV-MG. Fig.~\ref{Detailed} depicts the power stage of the ESS and the detailed block diagram of proposed control scheme. The battery storage unit is connected to a current source inverter (CSI) through bidirectional dc/dc converter in the power stage and the CSI is connected to the PCC of the MMG through a LC filter. In addition, there are single phase and three phase loads connected at the PCC. The modified reverse droop control scheme consists of a reactive power and a voltage compensator to improve the system power quality by regulating reactive powers.
\subsection{Proposed Modified Reverse Droop Control (MRDC)} The integration of the single-phase PV inverter and the three phase PV inverter increases the reactive power at the single-phase load ($Q_{L}$). This proposed controller decreases $Q_{L}$ by adjusting the reactive power of the ESS inverter $(Q_{s})$ to reduce the unbalance in the three-phase PCC voltage. $Q_{s}$ is calculated as follows: 
\begin{equation}
Q_{s}=2(\pm\sqrt{(P_{ESS})^{2}-(P_{s})^{2}}).
\label{PESS}
\end{equation}
where $P_{ESS}$ is the charge/discharge power of the ESS calculated in \eqref{PESS} by subtracting the single-phase load active power ($P_{L}$) from the maximum power ($P_{MPPT}$) generated by the single–phase PV array
\begin{equation}
P_{ESS}=P_{MPPT}-P_{L}.
\label{PESS}
\end{equation}

$P_{s}$ is shown in \eqref{PESS2} and it is the average active output power of the ESS inverter. This average output power is calculated using the instantaneous power ($p$) passed through a first-order low pass filter (LPF):
\begin{equation}
P_{s}=\frac{1}{2}.(v_{sd}.i_{sd}).
\label{PESS2}
\end{equation}
where $v_{sdq}$ and $i_{sdq}$  are the output voltage ($v_{s}$)  and output current ($i_{s}$) of the ESS  in dq reference frame respectively. 
Note that $v_{sq}$ is zero when the single-phase output voltage is placed on the d-axis \cite{daxis,single}. The magnitude ($V_{M}$) and the angular frequency ($w_{M}$) of the output voltage ($v_{s}$) are detected by phase locked loop (PLL) in the measurement block. The MRDC control consist of the RPC and the voltage compensator. 

\begin{figure*}[t]
\begin{subfigure}{.49\textwidth}
  \centering
\includegraphics[width=\linewidth]{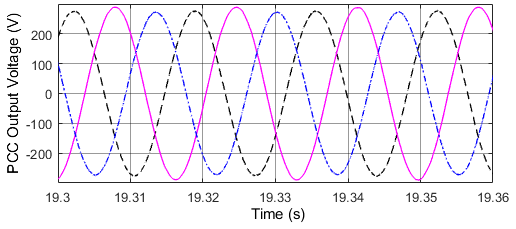} 
  \caption{}
\label{6ax}
\end{subfigure}
\begin{subfigure}{.49\textwidth}
  \centering
  \includegraphics[width=\linewidth]{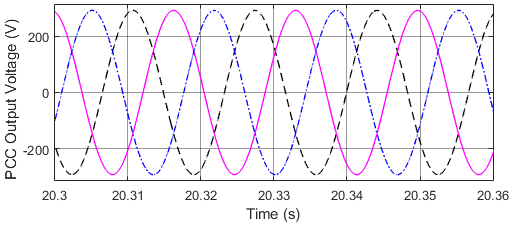}  
  \caption{}
\label{6bx}
\end{subfigure}
\caption{Output voltage of PCC before (a) and (b) after implementing RPCA.}
\label{RPCAx}
\vspace{-8pt}
\end{figure*}

\begin{figure*}[t]
\begin{subfigure}{.49\textwidth}
  \centering
\includegraphics[width=\linewidth]{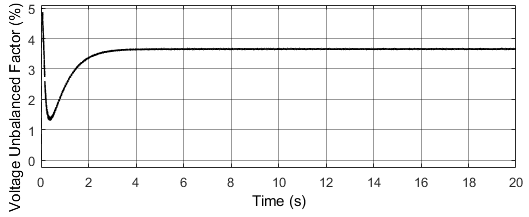} 
  \caption{}
\label{VUFa}
\end{subfigure}
\begin{subfigure}{.49\textwidth}
  \centering
  \includegraphics[width=\linewidth]{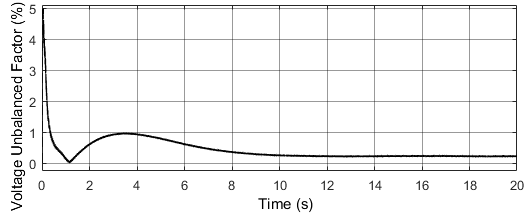}  
  \caption{}
\label{VUFb}
\end{subfigure}
\caption{VUF of PCC voltage before (a) and (b) after implementing RPCA.}
\label{VUF}
\vspace{-8pt}
\end{figure*}

\subsubsection{Reactive Power Compensator (RPC)} 
The proposed RPC compensates for the reactive powers of the MMG by adjusting the value of the output voltage and frequency of the ESS. Initially, $Q_{s}$ is produced by varying $P_{s}$. The reference active power ($P_{ref}$) of the ESS using droop control characteristic is defined as:

\begin{equation}
P_{ref}= P_{s}+(w_{ref}-w_{M})\frac{1}{n_{r}}
\label{x}
\end{equation}
where $w_{ref}$ is the angular frequency reference and $n_{r}$ is the active power reverse droop coefficient. The reference reactive power ($Q_{ref}$) of the ESS is expressed as: 
\begin{equation}
Q_{ref}= Q_{s}-(V_{ref}-V_{M})\frac{1}{m_{r}}-Q_{vcc} 
\label{x}
\end{equation}
where $V_{ref}$ is the output voltage reference of the single-phase PV inverter, $m_{r}$ is the reactive power reverse droop coefficient and $Q_{vcc}$ is the voltage compensation controller reference. It is worth mentioning that the reactive power of the single-phase PV inverter can be set equal by $Q_{ref}$ since the reactive power of single-phase load is decreased.
\begin{figure}
\centering
\includegraphics[width=\linewidth]{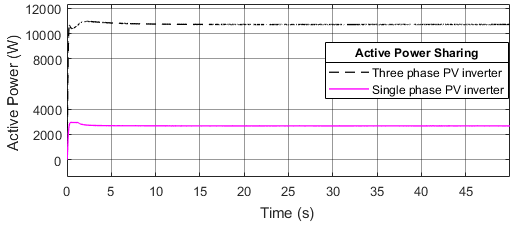}
\caption{Active Power Sharing Performance between PV-DGs.}
\label{RPSAx}
\vspace{-11pt}
\end{figure}

\subsubsection{Voltage Compensator} 
 This controller is also applied in the MRDC to balance the output voltage of the MMG. The reference of the voltage compensator ($Q_{vcc}$) is shown in \eqref{ql} as:
 \begin{equation}
\vspace{-1.75pt}
 Q_{vcc}= (V_{a}-V_{b})\frac{1}{d_{r}}+ (V_{c}-V_{a})\frac{1}{c_{r}} 
\label{ql}
\vspace{-1pt}
\end{equation}
where $V_{a}$, $V_{b}$ and $V_{c}$ are the phase voltage magnitudes from the three-phase PCC voltage ($V_{pcc}$) and $c_{r}$  and $d_{r}$ are voltage regulation coefficients. Note that $V_{a}$ is fixed since it is the output voltage of the single phase PV inverter. Therefore $V_{a}$ acts as the reference voltage and is compared with $V_{b}$ and $V_{c}$ to compensate for $V_{pcc}$. Finally, the reference currents ($i_{d}^{ref}$ and $i_{q}^{ref}$) in dq frame are calculated as:

\begin{align} 
\begin{split} 
i_{d}^{ref}= P_{ref}/v_{sd} \\ i_{q}^{ref}= Q_{ref}/v_{sd}.
\end{split}
\label{RSPA1}
\end{align}
\subsection{PR Current Controller}
The PR current controller regulates the inductance current. $i_{d}^{ref}$ and $i_{q}^{ref}$ in dq frame are transformed to the single phase output current reference ($i_{ref}$) of the ESS.  Then $i_{ref}$ is compared with measured instantaneous inductance current ($i_{L}$) of the ESS and $i_{L}$ is regulated by the PR current controller to produce single phase current reference. This reference is applied to pulse width modulator (PWM) block to control the switches of the CSI.


\begin{figure}[t]
\begin{subfigure}{.48\textwidth}
  \centering
\includegraphics[width=\linewidth]{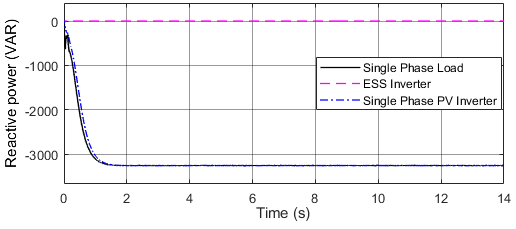} 
 \caption{}
\end{subfigure}
\begin{subfigure}{.48\textwidth}
  \centering
  \includegraphics[width=\linewidth]{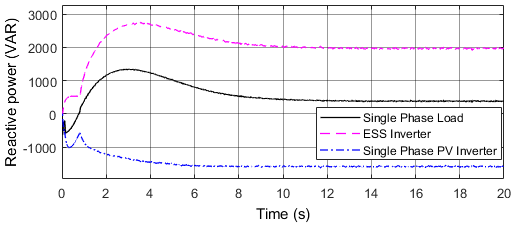}  
  \caption{}
  \label{PRAbx}
\end{subfigure}
\caption{Reactive powers (a) without and (b) with ESS strategy}
\label{PRAx}
\vspace{-13pt}
\end{figure} 
\section{Simulation and experimental validation} 
The proposed modified reverse droop control for an ESS in Fig.~\ref{Detailed} is implemented in the islanded MMG test system, as shown in Fig.~\ref{Simplified} to test unbalanced voltage quality in MATLAB/Simulink and real-time simulation using the Opal-RT OP5600 real-time simulator. The MMG system consists of a single ESS connected in parallel to the single-phase DG- PV system and the single phase load. Furthermore, the single-phase PV system is connected to one phase of the three-phase PV system considering different rated PV-DG power capacities at the PCC terminal to create unbalance voltage. Parameters of the system for the proposed controller is given in Table I. 

 \subsection{Simulation verification}  
 This  section  presents  the  simulation  results  of  the  test system as shown in Fig.~\ref{Simplified} using MATLAB/ Simulink.
\subsubsection{Performance of the modified reverse droop control (MRDC)} 
Fig.~\ref{6ax} and Fig.~\ref{VUFa} show the performance of the unbalanced MG before implementing the reactive power compensator. As seen in Fig.~\ref{6ax}, PCC output voltage is unbalanced and the voltage unbalance factor (VUF) of the $V_{pcc}$ is 3.6\% as shown in Fig.~\ref{VUFa}. After the modified reverse droop control is adopted. It can be seen that the three-phase PCC voltage quality is improved at the PCC terminal as shown in Fig.~\ref{6bx}. In order to demonstrate unbalance compensation more clearly, the VUF of the PCC terminal is decreased from 3.6\% to 0.25\% as seen in Fig.~\ref{VUFb} after compensation.


\subsubsection{Performance of the load power sharing} 

The connection of the single-phase PV systems and the loads to the three-phase PV systems changes active power of the single-phase PV inverter, therefore proportional active power sharing among the single-phase and the three-phase inverters is affected adversely. As a result of the reactive power compensation of the single-phase load, sharing of active power appropriately among different rated three-phase and single-phase PV-DG units is depicted in Fig.~\ref{RPSAx}. In addition, the reactive power of the three-phase inverter (10500 W) is 4 times reactive power of the single-phase PV inverter (2600 W) considering DG rated power capacities in Table I.  

\begin{table}
\caption{Simulation parameters}
\vspace{-2pt}
\begin{center}
\label{table}
\begin{tabular}{|c|c|}
\hline
\textbf{ESS System Parameter}   & \textbf{Value} \\ \hline
Switching frequency   & $F_{sw}$= 10 kHz \\ \hline
LC filter  &     $L$=1.5 mH $C$= 100 µF\\  \hline
DC link voltage & $v_{ESS,dc}^{ref}$= 300 V\\  \hline
Nominal voltage    &  $V_{ESS}^{ref}$= 120 rms V\\ \hline
System frequency  & $f_{ESS}^{ref}$= 60 rad/s \\ \hline
\textbf{MRDC Parameter} & \textbf{RPC/Voltage Compensator Value} \\ \hline
$n_{r}$, $m_{r}$                            & 0.67e-1, -0.8e-2 \\ \hline
$d_{r}$, $c_{r}$                          & -25e-6, -2.85e-4 \\ \hline
\textbf{PV and Load Parameter} & \textbf{ Power	Value} \\ \hline
Single-phase DG-PV Power &       $P_{PV}^{\phi}$ =3 kW  \\ \hline
Three-phase DG-PV Power &       $P_{PV}^{3\phi}$=12 kW \\ \hline
Single-phase PV load      &    20 $\Omega$           \\ \hline
\textbf{PR Control Parameter} & \textbf{Coefficient Value} \\ \hline
$k_{pI}$, $w_{cI}$                           &     40,   1             \\\hline
$k_{rIk}$                           &   1000(k = 1), 50(h = 3)          \\ \hline
\end{tabular}
\vspace{-15pt}
\end{center}
\end{table}

\begin{figure}
\centering
\includegraphics[width=\linewidth]{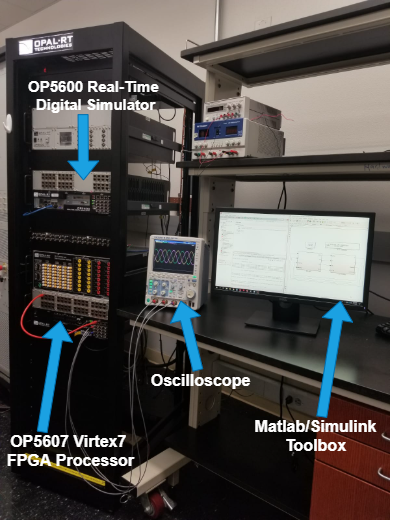}
\caption{Experimental setup for verification of the proposed control strategy.}
\label{opal}
\vspace{-11pt}
\end{figure}

\subsubsection{Performance of the reactive powers regulation} 

Fig.~\ref{PRAx} depicts the performance of the voltage compensator applied to an ESS. The reactive power of the single-phase load is minimized while the reactive output power ($Q_{s}$) of the ESS is increased to reduce the unbalanced voltage as shown in Fig.~\ref{PRAbx}.

\begin{figure*}[t]
\begin{subfigure}{.49\textwidth}
  \centering
\includegraphics[width=\linewidth]{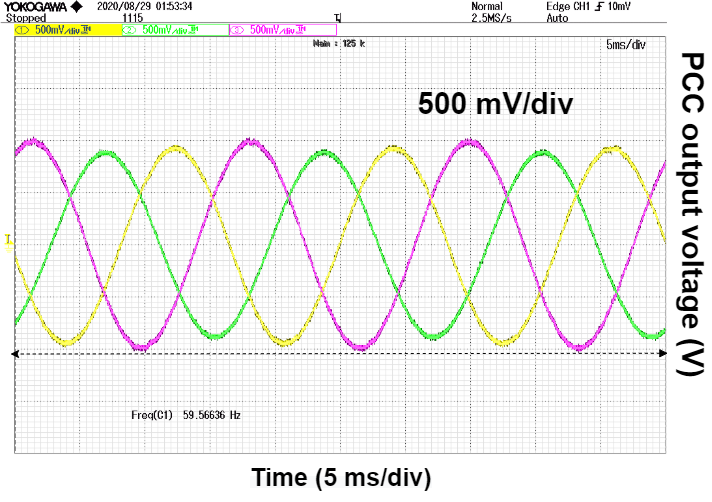} 
  \caption{}
\label{6a}
\end{subfigure}
\begin{subfigure}{.49\textwidth}
  \centering
  \includegraphics[width=\linewidth]{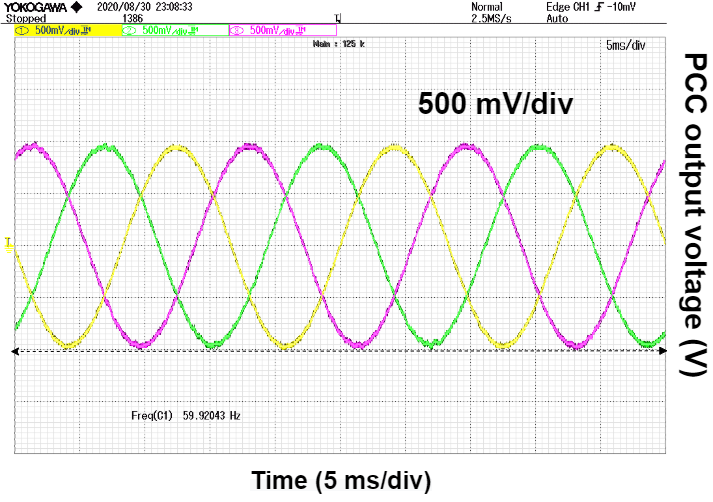}  
  \caption{}
\label{6b}
\end{subfigure}
\caption{Output voltage of PCC before (a) and (b) after implementing RPCA.}
\label{RPCA}
\vspace{-8pt}
\end{figure*}

\begin{figure}
\centering
\includegraphics[width=\linewidth]{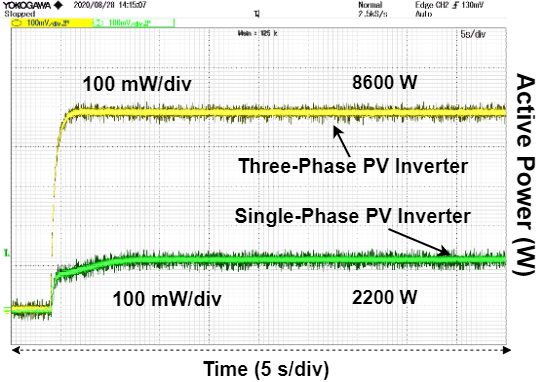}
\caption{Active Power Sharing Performance between PV-DGs.}
\label{RPSA}
\vspace{-11pt}
\end{figure}

\begin{figure}
\begin{subfigure}{.47\textwidth}
  \centering
\includegraphics[width=\linewidth]{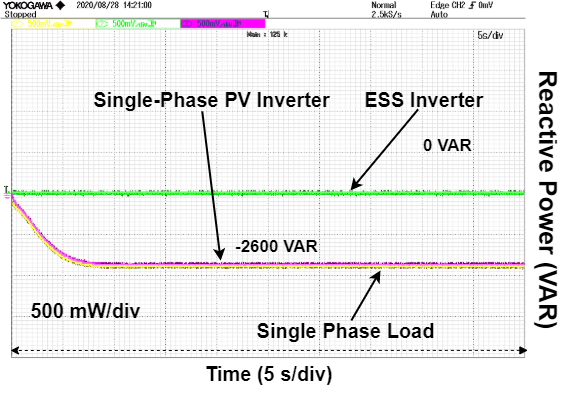} 
 \caption{}
\end{subfigure}
\begin{subfigure}{.47\textwidth}
  \centering
  \includegraphics[width=\linewidth]{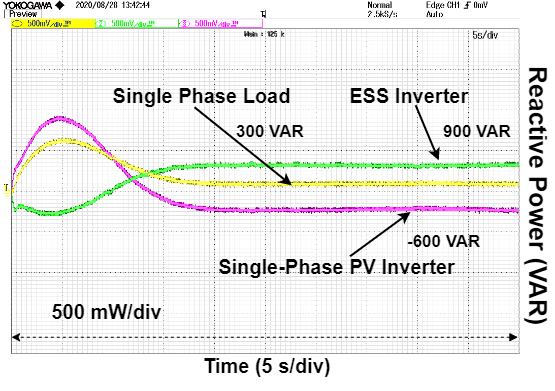}  
  \caption{}
  \label{PRAb}
\end{subfigure}
\caption{Reactive powers (a) without and (b) with ESS strategy}
\label{PRA}
\vspace{-8pt}
\end{figure} 

\subsection{Experimental verification}  
This section demonstrates the results of the real-time simulation of the proposed reversed droop control using Opal-RT OP5600 real-time simulator as shown in Fig.~\ref{opal}. The test system is simulated in two distinct models; an FPGA model and a CPU model on the Opal-RT simulator and the models are created in Matlab/Simulink SimScape Power System Toolbox. The power stage including the DGs, inverters, and loads is located on the OP5607 Virtex7 FPGA processor board for real-time power electronics simulation. The FPGA board runs at a smaller sampling time Ts = 0.5 µs and is connected to the digital and analog IO expansion units.  The proposed control method is implemented in the CPU model with a larger sampling time Ts = 15 µs. The CPU model is a classical OP5600 RT-LAB model and used to select the electrical model to be run on FPGA.

The real-time simulation results are also shown in Fig.~\ref{RPCA}-\ref{PRA} in order to verify the effectiveness of the proposed method.  It  can  be  observed  that  the real-time simulation results also demonstrate a noticeable improvement in the power quality. 

\section{conclusion}
This paper presents a coordinated control of an ESS in parallel to a single-phase PV inverter and the single-phase load for voltage unbalance mitigation in MMG. A novel modified reverse droop control scheme is implemented to improve the voltage quality at the PCC terminal. The control algorithm consisting of a reactive power compensator (RPC) and  a voltage compensator avoids complex calculations and requires only line voltage measurements. The real-time simulation results validate the effectiveness of this approach as the PCC voltage quality is improved by 70\%, while the load power sharing proportionally is achieved among the PV inverters. The reactive power of the single-phase PV inverter is changed by compensating reactive power of the single-phase load, which affects adversely reactive power sharing between single-phase and three-phase inverters. As the next step, we are working on the reactive power sharing between PV inverters.


%


\bibliographystyle{IEEEtran}
\bibliography{IEEEexample}

\begin{thebibliography}{10}
\providecommand{\url}[1]{#1}
\csname url@samestyle\endcsname
\providecommand{\newblock}{\relax}
\providecommand{\bibinfo}[2]{#2}
\providecommand{\BIBentrySTDinterwordspacing}{\spaceskip=0pt\relax}
\providecommand{\BIBentryALTinterwordstretchfactor}{4}
\providecommand{\BIBentryALTinterwordspacing}{\spaceskip=\fontdimen2\font plus
\BIBentryALTinterwordstretchfactor\fontdimen3\font minus
  \fontdimen4\font\relax}
\providecommand{\BIBforeignlanguage}[2]{{%
\expandafter\ifx\csname l@#1\endcsname\relax
\typeout{** WARNING: IEEEtran.bst: No hyphenation pattern has been}%
\typeout{** loaded for the language `#1'. Using the pattern for}%
\typeout{** the default language instead.}%
\else
\language=\csname l@#1\endcsname
\fi
#2}}
\providecommand{\BIBdecl}{\relax}
\BIBdecl

\bibitem{HMG2}
F.~{Nejabatkhah}, Y.~W. {Li}, and H.~{Tian}, ``Power quality control of smart
  hybrid ac/dc microgrids: An overview,'' \emph{IEEE Access}, vol.~7, pp.
  52\,295--52\,318, 2019.

\bibitem{RPCSP3}
R.~P.~S. {Chandrasena}, F.~{Shahnia}, S.~{Rajakaruna}, and A.~{Ghosh},
  ``Operation and control of three phase microgrids consisting of single-phase
  ders,'' in \emph{2013 IEEE 8th International Conference on Industrial and
  Information Systems}, 2013, pp. 599--604.

\bibitem{ESShybrid}
S.~{Resch} and M.~{Luther}, ``The combination of single- and three-phase
  inverters into a hybrid energy storage system,'' in \emph{2020 IEEE PES
  Innovative Smart Grid Technologies Europe (ISGT-Europe)}, 2020, pp. 965--969.

\bibitem{KHAN2021}
A.~A. Khan, O.~A. Beg, M.~Alamaniotis, and S.~Ahmed, ``Intelligent anomaly
  identification in cyber-physical inverter-based systems,'' \emph{Electric
  Power Systems Research}, vol. 193, p. 107024, 2021.

\bibitem{apec}
E.~{Pritchard}, L.~{Mackey}, D.~{Zhu}, D.~{Gregory}, and G.~{Norris}, ``Modular
  electric generator rapid deployment dc microgrid,'' in \emph{2017 IEEE Second
  International Conference on DC Microgrids (ICDCM)}, 2017, pp. 106--110.

\bibitem{akdogan}
M.~E. Akdogan \emph{et~al.}, ``Advanced power sharing scheme under unbalanced
  and nonlinear loads in an islanding microgrid,'' Ph.D. dissertation, 2017.

\bibitem{IET}
A.~S. {Vijay}, S.~{Doolla}, and M.~C. {Chandorkar}, ``Unbalance mitigation
  strategies in microgrids,'' \emph{IET Power Electronics}, vol.~13, no.~9, pp.
  1687--1710, 2020.

\bibitem{MMG}
C.~Wang, P.~Yang, R.~Chen, S.~Cheng, T.~Tian, and X.~Li, ``Hierarchical voltage
  imbalance control for single-/three-phase hybrid multimicrogrid,'' \emph{IET
  Generation, Transmission \& Distribution}, vol.~13, no.~18, pp. 4233--4241,
  2019.

\bibitem{HMG1}
J.~{Zhou}, Y.~{Xu}, H.~{Sun}, Y.~{Li}, and M.~{Chow}, ``Distributed power
  management for networked ac–dc microgrids with unbalanced microgrids,''
  \emph{IEEE Transactions on Industrial Informatics}, vol.~16, no.~3, pp.
  1655--1667, 2020.

\bibitem{mypaper}
M.~E. {Akdogan} and S.~{Ahmed}, ``Control hardware-in-the-loop for voltage
  controlled inverters with unbalanced and non-linear loads in stand-alone
  photovoltaic (pv) islanded microgrids,'' in \emph{2020 IEEE Energy Conversion
  Congress and Exposition (ECCE)}, 2020, pp. 2431--2438.

\bibitem{apf}
M.~M. {Hashempour}, M.~{Savaghebi}, J.~C. {Vasquez}, and J.~M. {Guerrero},
  ``Voltage unbalance and harmonic compensation in microgrids by cooperation of
  distributed generators and active power filters,'' in \emph{2016 7th Power
  Electronics and Drive Systems Technologies Conference (PEDSTC)}, 2016, pp.
  646--651.

\bibitem{UPQ}
D.~{Graovac}, V.~{Katic}, and A.~{Rufer}, ``Power quality problems compensation
  with universal power quality conditioning system,'' \emph{IEEE Transactions
  on Power Delivery}, vol.~22, no.~2, pp. 968--976, 2007.

\bibitem{af}
S.~{Shukla}, B.~{Singh}, and S.~{Mishra}, ``Implementation of empirical mode
  decomposition for shunt active filter,'' in \emph{2015 IEEE IAS Joint
  Industrial and Commercial Power Systems / Petroleum and Chemical Industry
  Conference (ICPSPCIC)}, 2015, pp. 174--181.

\bibitem{sequence}
P.~{Mishra}, A.~K. {Pradhan}, and P.~{Bajpai}, ``Voltage control of pv inverter
  connected to unbalanced distribution system,'' \emph{IET Renewable Power
  Generation}, vol.~13, no.~9, pp. 1587--1594, 2019.

\bibitem{03}
C.~{Wang}, P.~{Yang}, C.~{Ye}, Y.~{Wang}, and Z.~{Xu}, ``Voltage control
  strategy for three/single phase hybrid multimicrogrid,'' \emph{IEEE
  Transactions on Energy Conversion}, vol.~31, no.~4, pp. 1498--1509, 2016.

\bibitem{Multiobjective}
D.~I. {Brandao}, W.~M. {Ferreira}, A.~M.~S. {Alonso}, E.~{Tedeschi}, and F.~P.
  {Marafão}, ``Optimal multiobjective control of low-voltage ac microgrids:
  Power flow regulation and compensation of reactive power and unbalance,''
  \emph{IEEE Transactions on Smart Grid}, vol.~11, no.~2, pp. 1239--1252, 2020.

\bibitem{RPCSP1}
M.~{Zeraati}, M.~E.~H. {Golshan}, and J.~M. {Guerrero}, ``Voltage quality
  improvement in low voltage distribution networks using reactive power
  capability of single-phase pv inverters,'' \emph{IEEE Transactions on Smart
  Grid}, vol.~10, no.~5, pp. 5057--5065, 2019.

\bibitem{RPCSP2}
F.~{Nejabatkhah} and Y.~W. {Li}, ``Flexible unbalanced compensation of
  three-phase distribution system using single-phase distributed generation
  inverters,'' \emph{IEEE Transactions on Smart Grid}, vol.~10, no.~2, pp.
  1845--1857, 2019.

\bibitem{ESS1}
M.~{Zeraati}, M.~E. {Hamedani Golshan}, and J.~M. {Guerrero}, ``Distributed
  control of battery energy storage systems for voltage regulation in
  distribution networks with high pv penetration,'' \emph{IEEE Transactions on
  Smart Grid}, vol.~9, no.~4, pp. 3582--3593, 2018.

\bibitem{ESS2}
K.~H. {Chua}, Y.~S. {Lim}, P.~{Taylor}, S.~{Morris}, and J.~{Wong}, ``Energy
  storage system for mitigating voltage unbalance on low-voltage networks with
  photovoltaic systems,'' \emph{IEEE Transactions on Power Delivery}, vol.~27,
  no.~4, pp. 1783--1790, 2012.

\bibitem{chinmay}
C.~{Shah} and R.~{Wies}, ``Algorithms for optimal power flow in isolated
  distribution networks using different battery energy storage models,'' in
  \emph{2020 IEEE Power Energy Society Innovative Smart Grid Technologies
  Conference (ISGT)}, 2020, pp. 1--5.

\bibitem{Droop1}
E.~{Espina}, R.~{Cárdenas-Dobson}, M.~{Espinoza-B.}, C.~{Burgos-Mellado}, and
  D.~{Sáez}, ``Cooperative regulation of imbalances in three-phase four-wire
  microgrids using single-phase droop control and secondary control
  algorithms,'' \emph{IEEE Transactions on Power Electronics}, vol.~35, no.~2,
  pp. 1978--1992, 2020.

\bibitem{Droop2}
{Dan Wu}, {Fen Tang}, J.~C. {Vasquez}, and J.~M. {Guerrero}, ``Control and
  analysis of droop and reverse droop controllers for distributed
  generations,'' in \emph{2014 IEEE 11th International Multi-Conference on
  Systems, Signals Devices (SSD14)}, 2014, pp. 1--5.

\bibitem{Droop3}
A.~{Villa}, F.~{Belloni}, R.~{Chiumeo}, and C.~{Gandolfi}, ``Conventional and
  reverse droop control in islanded microgrid: Simulation and experimental
  test,'' in \emph{2016 International Symposium on Power Electronics,
  Electrical Drives, Automation and Motion (SPEEDAM)}, 2016, pp. 288--294.

\bibitem{Droop4}
P.~P. {Vergara}, J.~C. {Lopez}, M.~J. {Rider}, and L.~C.~P. {da Silva},
  ``Optimal operation of unbalanced three-phase islanded droop-based
  microgrids,'' \emph{IEEE Transactions on Smart Grid}, vol.~10, no.~1, pp.
  928--940, 2019.

\bibitem{Droop5}
S.~Y. {Mousazadeh Mousavi}, A.~{Jalilian}, M.~{Savaghebi}, and J.~M.
  {Guerrero}, ``Autonomous control of current- and voltage-controlled dg
  interface inverters for reactive power sharing and harmonics compensation in
  islanded microgrids,'' \emph{IEEE Transactions on Power Electronics},
  vol.~33, no.~11, pp. 9375--9386, 2018.

\bibitem{ESSdroop1}
C.~{Jamroen}, A.~{Pannawan}, and S.~{Sirisukprasert}, ``Battery energy storage
  system control for voltage regulation in microgrid with high penetration of
  pv generation,'' in \emph{2018 53rd International Universities Power
  Engineering Conference (UPEC)}, 2018, pp. 1--6.

\bibitem{ESSdroop2}
N.~{Mahmud}, A.~{Zahedi}, and M.~S. {Rahman}, ``An event-triggered distributed
  coordinated voltage control strategy for large grid-tied pv system with
  battery energy storage,'' in \emph{2017 Australasian Universities Power
  Engineering Conference (AUPEC)}, 2017, pp. 1--6.

\bibitem{daxis}
N.~Akel, M.~Pahlevaninezhad, and P.~Jain, ``A dq rotating frame reactive power
  controller for single-phase bi-directional converters,'' in \emph{2014 IEEE
  36th International Telecommunications Energy Conference (INTELEC)}.\hskip 1em
  plus 0.5em minus 0.4em\relax IEEE, 2014, pp. 1--5.

\bibitem{single}
S.~{Peng}, A.~{Luo}, Z.~{Lv}, J.~{Wu}, and L.~{Yu}, ``Power control for
  single-phase microgrid based on the pq thoery,'' in \emph{2011 6th IEEE
  Conference on Industrial Electronics and Applications}, 2011, pp. 1274--1277.

\end{thebibliography}
\end{document}